\documentstyle[11pt,aaspp4]{article}

\def\l{$\ell$}
\def\m{$m$}
\def\k{$k$}
\def\etal{et~al.\,}
\def\s2n{signal to noise}
\def\agy{asteroseismology }
\def\agl{asteroseismological }
\def\g29{G29--38}
\def\Msun{{\rm M}_\odot}                  
\def\Mstar{{{\rm M}\lower.5ex\hbox{$\star$}}}
\def\mhz{${\rm \mu Hz}$}

\def\ylm{$Y_\ell^m(\theta,\phi)$}
\def\dpi{$\Delta \Pi$}
\def\o-c{$O-C$}
\def\pg1159{PG~1159--035}
\def\gd358{GD~358}

\def\ie{i.e.\ }

\def\pdot{$\dot{{\rm P}}$}

\def\191g{G191--16}
\def\38g{G38--29}

\slugcomment{To appear in ApJ., 1Mar98}

\lefthead{Kleinman \etal}
\righthead{DA Pulsators}

\begin{document}

\title{Understanding the Cool DA White Dwarf Pulsator, G29--38}
\author{S.~J.~Kleinman,\altaffilmark{1,2,3}
R.~E.~Nather,\altaffilmark{2}
D.~E.~Winget,\altaffilmark{2,4}
J.~C.~Clemens,\altaffilmark{5,6,4}
P.~A.~Bradley,\altaffilmark{7}
A.~Kanaan,\altaffilmark{8,9}
J.~L.~Provencal,\altaffilmark{10,11}
C.~F.~Claver,\altaffilmark{12}
T.~K.~Watson,\altaffilmark{13}
K.~Yanagida,\altaffilmark{2,14}
A.~Nitta,\altaffilmark{2}
J.~S.~Dixson,\altaffilmark{15}
M.~A.~Wood,\altaffilmark{16,17}
A.~D.~Grauer,\altaffilmark{18,19}
B.~P.~Hine,\altaffilmark{20}
G.~Fontaine,\altaffilmark{21,4}
James~Liebert,\altaffilmark{22}
D.~J.~Sullivan,\altaffilmark{23}
D.~T.~Wickramasinghe,\altaffilmark{24}
N.~Achilleos.\altaffilmark{24}
T.~M.~K.~Marar,\altaffilmark{25}
S.~Seetha,\altaffilmark{25}
B.~N.~Ashoka,\altaffilmark{25}
E.~Mei\v{s}tas,\altaffilmark{26,27}
E.~M.~Leibowitz,\altaffilmark{28}
P.~Moskalik,\altaffilmark{29}
J.~Krzesi\'nski,\altaffilmark{30}
J.-E.~Solheim,\altaffilmark{31,27}
A.~Bruvold,\altaffilmark{31}
D.~O'Donoghue,\altaffilmark{32}
D.~W.~Kurtz,\altaffilmark{32}
B.~Warner,\altaffilmark{32}
Peter~Martinez,\altaffilmark{32}
G.~Vauclair,\altaffilmark{33}
N.~Dolez,\altaffilmark{33}
M.~Chevreton,\altaffilmark{34}
M.~A.~Barstow,\altaffilmark{35,36,9}
S.~O.~Kepler,\altaffilmark{8}
O.~Giovannini,\altaffilmark{8,37}
T.~Augusteijn,\altaffilmark{38}
C.~J.~Hansen,\altaffilmark{39}
\and
S.~D.~Kawaler\altaffilmark{13}}

\altaffiltext{1}{New Jersey Institute of Technology, Big Bear Solar Observatory, 40386
North Shore Lane, Big Bear City, CA 92314: sjk@begonias.bbso.njit.edu}
\altaffiltext{2}{Astronomy Department, University of Texas, Austin TX 78712}
\altaffiltext{3}{Guest Observer, Mount Stromlo and Siding Spring Observatory,
N.S.W., Australia}
\altaffiltext{4}{Visiting Astronomer, Canada-France-Hawaii Telescope,
operated by the National Research Council of Canada, Centre National de la
Recherche Scientifique de France, and the University of Hawaii}
\altaffiltext{5}{California Institute of Technology, Pasadena CA 91125}
\altaffiltext{6}{Fairchild Fellow}
\altaffiltext{7}{X-2, MS B-220, Los Alamos National
Laboratory, Los Alamos, NM 87545}
\altaffiltext{8}{Instituto de F\'{\i}sica, Universidade Federal do Rio Grande
do Sul, 91501-970 Porto Alegre-RS, Brazil}
\altaffiltext{9}{Guest Observer, Isaac Newton Telescope, Roque de los
Muchachos, La Palma, Canaries}
\altaffiltext{10}{University of Delaware, Physics and
Astronomy Department, Sharp Laboratory, Newark DE, 19716}
\altaffiltext{11}{Guest Observer, Cerro Tololo Inter-American Observatory,
Chile}
\altaffiltext{12}{NOAO, 950 N. Cherry Ave., Tucson AZ 85726}
\altaffiltext{13}{Department of Physics and Astronomy, Iowa State University, 
Ames IA 50211}
\altaffiltext{14}{Current postal address: 5-35-11 Hong\=odai, Sakae-ku
Yokohama 247, Japan}
\altaffiltext{15}{M/S ADVP3, DSC Communications Corp., 1000 Coit Rd Plano, TX 75075}
\altaffiltext{16}{Department of Physics and Space Sciences, Florida Institute
of Technology, 150 West University Boulevard, Melbourne FL 32901}
\altaffiltext{17}{Guest Observer, Institute for Astronomy, Honolulu, HI}
\altaffiltext{18}{Department of Physics and Astronomy, University of
Arkansas, Little Rock, AR, 72204}
\altaffiltext{19}{Visiting Astronomer, Kitt Peak National Observatory}
\altaffiltext{20}{NASA Ames Research Center, MS 269-3, Moffett Field, CA,
94035}
\altaffiltext{21}{Department de Physique, Universit\'e de Montr\'eal, C.P.
6128, Montreal, PQ, H3C 3J7 Canada}
\altaffiltext{22}{Steward Observatory, University of Arizona, Tucson AZ
85721}
\altaffiltext{23}{Department of Physics, Victoria University, Box 600,
Wellington, New Zealand}
\altaffiltext{24}{Department of Mathematics, Australia National University, 
Canberra, Australia}
\altaffiltext{25}{Indian Space Research Organization, Bangalore 560 017, India}
\altaffiltext{26}{Institute of Theoretical Physics and Astronomy, Go\v{s}tauto
12, Vilnius 2600, Lithuania}
\altaffiltext{27}{Guest Observer, Maidanak Observatory,Uzbekistan}
\altaffiltext{28}{Department of Physics and Astronomy, University of Tel
Aviv, Ramat Aviv, Tel Aviv 69978, Israel}
\altaffiltext{29}{Copernicus Astronomical Center, Warsaw, Poland}
\altaffiltext{30}{Mount Suhora Observatory, Cracow Pedagogical University,
ul. Podchor\c{a}\.{z}ych 2, 30-084 Cracow, Poland}
\altaffiltext{31}{Institutt for Matematiske Realfag, Universitet i Tromso,
9000 Tromso, Norway}
\altaffiltext{32}{Department of Astronomy, University of Cape Town,
Rondebosch 7700, Cape Province, South Africa}
\altaffiltext{33}{Observatoire Midi-Pyrenees, 14 Avenue E. Belin, 31400
Toulouse, France}
\altaffiltext{34}{Observatoire de Paris-Meudon, F-92195 Meudon, Principal
Cedex, France}
\altaffiltext{35}{Department of Physics and Astronomy, University of
Leicester, Leicester LE1 7RH, UK}
\altaffiltext{36}{PPARC Advanced Fellow}
\altaffiltext{37}{Visiting Astronomer, Laboratorio Nacional de Astrofisica,
CNPq, Brazil}
\altaffiltext{38}{European Southern Observatory Casilla 19001 Santiago~19
Chile}
\altaffiltext{39}{JILA, University of Colorado, Box 440, Boulder, CO 80309}

\setcounter{footnote}{0}

\begin{abstract}

The white dwarfs are promising laboratories for the study of
cosmochronology and stellar evolution.  Through observations of the
pulsating white dwarfs, we can measure their internal structures
and compositions, critical to understanding post main sequence
evolution, along with their cooling rates, allowing us to 
calibrate their ages directly.  The most important set of white dwarf
variables to measure are the oldest of the pulsators, the cool DAVs, which
have not previously been explored through asteroseismology due to their
complexity and instability.  Through a time-series photometry data set
spanning ten years, we explore the pulsation spectrum of the cool DAV,
G29--38 and find an underlying structure of 19 (not including multiplet
components) normal-mode, probably $\ell=1$
pulsations amidst an abundance of time variability and linear combination
modes.  Modelling results are incomplete, but we suggest possible starting
directions and discuss probable values for the 
stellar mass and hydrogen layer size. For the first time, we have made
sense out of the complicated power spectra of a large-amplitude DA
pulsator. We have shown its seemingly erratic set of observed frequencies
can be understood in terms of a recurring set of normal-mode pulsations and
their linear combinations.  With this result, we have opened the interior
secrets of the DAVs to future \agl modelling, thereby joining the rest
of the known white dwarf pulsators.

\end{abstract}


\section{Introduction}

Due to their status as the end-products of most stellar evolution paths, the
white dwarfs are among the oldest stars in the galaxy and therefore offer
important clues about the universe around us.
Studying their interiors will provide solid endpoints for
stellar evolution, providing insights into nuclear reaction rates, mass
loss mechanisms, and basic physical properties of matter in a variety of
extreme conditions.  Dating the oldest of these stars provides natural
constraints on the age of our Galaxy, including possible independent
measurements of the ages of the disk, the halo, and open and globular
clusters.

The key to exploiting the potential of the white dwarfs is buried
beneath their relatively thin surface layers: we must discover their
internal structure and composition to make them useful. Measuring the
ages of the
oldest white dwarfs, which have spent only a small fraction of their
entire existence {\it off} the white dwarf cooling track, is as simple
as measuring their cooling rates.  Theoretical cooling rate models depend
on the mass and structure of the white dwarf. Asteroseismology can both
determine these physical parameters of a star and provide a
means for {\it directly} measuring a star's cooling rate.

The DAV (or ZZ~Ceti) white dwarfs, with hydrogen-dominated
spectra, are the coolest ($\approx 13,000$K) and oldest of the four
known classes of white dwarf nonradial g-mode pulsators.\footnote{The
others three classes are the PNNVs, the DOVs, and the DBVs.  See
\cite{bro94}, for example, for a review of asteroseismology in a
broader context.} As such, they are the most critical in answering the
age question.  Unfortunately, they have also proven the most difficult to
understand.

The techniques of \agy work best when provided with an abundance of
stable modes of oscillation. The first white dwarf successfully
analyzed, the DOV, \pg1159, (\cite{pg1159}) provided over 100 such
modes.  The best analyzed DBV, \gd358 (\cite{gd358}), obliged us
similarly.  Until now, however, the DAVs were not so willing to
provide us with the necessary quantity of modes for easy analysis.

The DAVs exhibit distinct trends with temperature:  the hotter stars
have lower-amplitude, shorter-period pulsations and the cooler ones
have large-amplitude, longer-period pulsations. The hot DAVs have very
few modes while the cooler DAVs have many, but most modes are unstable and
seem to come and go without forming any obvious patterns of behavior.

In his dissertation, Clemens (1993, 1994) found additional systematic
properties in the hotter DAVs by looking at an ensemble of individual
pulsators, thereby establishing a technique for applying
asteroseismology on these objects which were thought to have too few
modes for analysis.  He was able to determine the individual properties
of each of the hot DAVs he analyzed by seeing how it fitted in with
group properties he discovered.  In each star he found only a few
modes, but when added together, they formed a much larger, coherent set
of \l=1 modes and a few \l=2 modes.\footnote{The surface deformations
on a pulsating white dwarf are usually modelled by the set of spherical
harmonics, \{\ylm\} and an additional radial-node index, \k. \l\ is the
total number of surface nodes and \m\ is the number of surface nodes
along a line of longitude.  Most observed modes are \l=1, with a
smattering of \l=2.  There have been no reliable identifications of an \l=3 (or
higher) mode in any white dwarf star.}  Clemens found a successive
series of \l=1 modes from \k=1 to \k=6 even though no one star had
modes at each value of \k. Each observed mode, however, was common to
more than one star;  when a star had an observed mode, it was always in
one of the discerned groups.

The implication of this result is that the overall structure
of the hot DAVs must be very similar. Comparing the set of observed modes to
theoretical models suggests the masses of the stars Clemens studied are quite near 
$0.6\Msun$ and their hydrogen envelopes are near $10^{-4}\Mstar$ (where $\Mstar$ is the
total mass of the star).  The
remarkable similarity of the hot DAVs supports the common
\agy credo that the pulsators are ``otherwise normal stars,'' although
the cool DAVs with their complex, variable power spectra remained,
for the time-being, an enigma.  The debate over ``thick'' versus ``thin''
Hydrogen layers, however, is far from over. See for example Fontaine \etal
(1994), Fontaine and Wesemael (1997), and recent modelling results from
Bradley (1997) for a recent summary.

Here, we present results from an extensive study of one of the cool DAVs,
\g29.  We find that although its power spectrum is not stable and changes
quite dramatically from one observing season to another (with smaller
changes within a season), we can still fit the observed set of
pulsation modes with a set of predominantly \l=1 normal-mode pulsations.
The key to this
result was two-part: 1) obtain many seasons of data to observe a larger 
set of available modes, and 2) accurately identify and separate the
linear-combinations from the more fundamental modes of oscillation.
The linear-combinations are observed periodicities whose frequencies
are sums and differences of those of other modes; we believe they arise
mainly from non-linear effects in the system.

There has been a great deal of previous interest in \g29. Its variability was
discovered by Shulov and Kopatskaya (1974) and confirmed by McGraw and
Robinson (1975). Winget \etal (1990) observed it in 1988 and found a
still unexplained phase variation of the dominant oscillation period,
but did not explain the bulk of the star's pulsation properties.
Zuckerman and Becklin (1987) reported a significant infrared excess in
the star at wavelengths longer than 2\micron, the source of which is
also still unknown, although orbiting dust is becoming the consensus
(\cite{zuc93}, \cite{koe97}).  Barnbaum \& Zuckerman (1992) report a possibly
periodic radial velocity variation with an unknown source.  Kleinman
\etal (1994) used the phase timings of  a stable pulsation mode at 284s
to place severe limits on what kinds of orbital companions could be
included in the system, showing the radial velocity variations cannot
be orbital in origin.  A recurring thread throughout most
of these works is the variability and complexity of \g29's power
spectra.  For the first time, we now propose a simple picture for
\g29's pulsations.

Our results, combined with Clemens's work on the hotter DAVs, place
all the DAVs alongside their other pulsating white dwarf cousins and
open this crucial class of variables for asteroseismological
analysis.  We show there is a pattern behind the complex variable
nature of these systems and, in the process, discover a
different kind of data set is needed to solve these stars.  A single
set of observations, over a single observing season, no matter how
well-resolved it is, will not suffice.  There must be well-sampled
observations over many observing seasons.

\section{Techniques}

The white dwarfs are particularly rewarding objects for
asteroseismology, the gains from which increase with the
number of identified modes.  (Modes are identified
by specifying the values of the three integers, \k, \l, and \m, along with
$\nu$ or $P$, the frequency or period.)  Cepheids pulsate in one or two
observed modes; the $\delta$ Scuti star with the most known modes has around
20 (\cite{bre95}). The white dwarfs, however, can pulsate in
hundreds of observable modes.  Each normal mode probes a slightly
different region of the stellar interior, so having 100 modes is like
having 100 probes, all going to different depths and locations in
the star's interior.

The advantages of white dwarfs as asteroseismological laboratories
quickly become their biggest disadvantage as well: they are very
complicated.  With so many modes active at the same time, we need
extended data sets to be able to resolve and identify closely-spaced
modes since the resolution
of the Fourier transform (FT), which we use to search for periodicities
in our reduced lightcurves, is proportional to the inverse of the
time duration of the lightcurve.  Even with a high-resolution FT, we can
still have problems resolving closely-spaced modes if the alias peaks
caused by observing gaps end up near real modes of the star.
The FT finds an inevitable ambiguity in cycle count of each measured
frequency due to these gaps.  For nightly observations from a
single-site, we therefore get alias peaks separated from the real ones by
integral multiples of one cycle ${\rm d}^{-1}$.  Unfortunately, this is 
near the typical white dwarf rotation rate which causes {\it real} modes to
be present, separated by $\approx 1{\rm d^{-1}}$ in frequency.

In order to eliminate these troublesome
$1{\rm d^{-1}}$ aliases, we strive to eliminate
the one day gap in our data.  To do this, we set up a network of
collaborating astronomers around the globe, all observing the same star
over the same time period with similar tools and observing techniques.
This network,  called the Whole Earth Telescope (WET: \cite{wet1}), has
been used quite successfully in the study of DO and DB variables.  With
the WET, we can obtain data uninterrupted by the daily rising of the
sun and hence produce vastly improved FTs, with few aliases surrounding
the real peaks.

Armed with a nearly alias-free transform, the goal of \agy is to match
each observed frequency with a unique value of \k, \l, and \m.   If a
full set of modes (say all the possible \m\ values for at least two values
of \l\ over a consecutive series of \k\ values) are present in the
star, the job is relatively easy.  If not, then clues must be taken
whenever available and pieced together for a consistent final picture.
The clues involve the spacings between modes of same \k\ and \l\, but
different \m, and those between modes of same \l\ and \m, but different
\k.

As the number of radial nodes (\k) increases, the frequency of a g-mode
decreases (there is less of a restoring force since the wavelength of
the oscillation decreases, meaning less mass is present to supply the 
gravitational restoring force).  In the asymptotic (high \k) limit, the
modes with same \l\ are equally spaced in period.  The periods $P$, of 
such modes are given by (see, for example \cite{5horse}):
\begin{equation}
P=\frac{k  \Delta \Pi}{\sqrt{\ell (\ell +1)}} + constant
\end{equation}
where \dpi\ is a constant related to the period spacing and the
additive constant is small.  \dpi\ itself is primarily a function of
the mass of the star and is truly constant only for stars of uniform
composition.  Adding compositionally stratified layers (white dwarf stars
have a very high gravity which separates and stratifies the constituent
elements), or any other radial discontinuity, to a model star makes the
value of \dpi\ different for each mode, although the mean remains a good
measure of the total stellar mass.  The deviations from uniformity of \dpi\
measure the layering present in the star.  This effect is called
mode-trapping as modes with nodes near the composition transition
boundaries will have their periods shifted slightly so the nodes correspond
more closely to the transition discontinuities (\cite{win81},
\cite{bras92}).

As long as the spherical harmonics are valid representations of the
observed surface distortions (they are as valid as our
assumption of spherical symmetry), their underlying symmetry implies
that the period of each mode depends only on \k, and \l, the total
number of surface nodal lines and not on how the nodes are
arranged on the surface (\ie, \m).  When the underlying symmetry is
broken, however, the observed periods become a function of \m\ as
well.  Rotation breaks the symmetry and splits each mode of a given
\k\ and \l\ into a multiplet of modes with $2\ell + 1$ components with
\m\ running from --\l\ to \l.

For slow rotation, the frequency difference ($\Delta \sigma$)
for each \m\ mode is given, to first order, by:

\begin{equation}
\Delta \sigma = m (1-C_{\ell,k}) \Omega
\end{equation}
where $\Omega$ is the constant stellar rotation frequency and
$C_{\ell,k}$ is, in general, a complicated function of the star's
density and modal displacements which in the asymptotic (large \k)
limit, approaches the value 1/\l(\l+1). Thus, modes of
the same \k\ and \l, but different \m\ will be uniformly spaced in
frequency and, barring radially differential rotation, all modes with the
same \l\ will have the same frequency spacings.  If radially
differential rotation is present, which changes the effective value of
$\Omega$ for each mode,
we will see a corresponding systematic pattern in the \m-splittings as
a function of \k.  Since the white dwarf rotation periods so far
measured with the WET are near one day, we expect the \l=1 \m-spacing
to be on the order of 6\mhz, much smaller than the typical period
spacing, \dpi, which is close to 150\mhz\ in the region of main power.
Since these rotational spacings are small, we expect to find
closely-spaced triplets for \l=1 modes and quintuplets for \l=2.

Having accurately isolated and measured the frequencies of the
pulsations, with a long timebase of observations we can actually observe
stellar evolution in progress.  The period of a mode changes slightly
as the star cools and/or contracts.  The contraction process (presumably
dominating only in the hottest of the pre-white dwarf pulsators, since
white dwarfs cool at essentially constant radius) 
decreases the period with time while cooling increases it.
Since the rates of change (or \pdot s) are very small (of order
$10^{-15} s/s$ for an average DAV), this is a very difficult measurement,
but an extremely important one.  Once we can measure the rate at
which a star is cooling (\cite{pg1159}, \cite{kep93}), we can empirically
calibrate the white dwarf cooling curve, and hence the luminosity
function (\cite{lie88}) to measure directly the age of the white dwarfs
(\cite{win87}).

\section{Observations}

The WET is the ideal tool for analysis of stars with stable pulsation spectra.
For most DAVs, however, it is not enough.  During any
given season of observations, no matter how well resolved the power
spectrum of any one star, only a few independent pulsation modes are
present; the rest of the frequencies found in the power spectrum are
linear combinations of existing modes.  If, however, we observe the same star over
many seasons, we can add a few new modes each year and slowly build
them up to see a larger set of possible modes.  This approach will work
only if there is a stable underlying pattern to the star's
pulsations: that is, if it picks and chooses a few modes each season
from a pre-defined, limited set of possibilities. We found this to
be the case with \g29.

Presented here are over 1100 hours of time-series photometry (we call the
technique {\it temporal spectroscopy}) on \g29,  representing the
results of three campaign observations (two WET runs and a double-site
venture between SAAO and McDonald) and many years of intense single-site
coverage.   They span the 10 years from 1985 to 1994 with data every
year except 1986 and 1987.  Rather than present the complete lengthy
table of observations here, which can be found in its entirety in
Kleinman (1995) and nearly complete in Kleinman \etal (1994), we list in
Table~\ref{tb:sitelist} only the sites and telescopes which have
contributed data, and in Table~\ref{tb:addjobs} the additional
observations used in this work that were not listed in the Kleinman
\etal (1994) paper.

\begin{deluxetable}{lr}
\tablecaption{\label{tb:sitelist}Data-providing sites.}


\makeatletter
\def\jnl@aj{AJ}
\ifx\revtex@jnl\jnl@aj\let\tablebreak=\nl\fi
\makeatother

\tablewidth{16pc}
\tablehead{
\colhead{Location}           & \colhead{Telescope}}

\startdata
CTIO & 1.5m \nl
Itajuba, LNA & 1.6m \nl
KPNO & 1.3m \nl
La Palma (INT)& 2.5m \nl
Maidanak & 1.0m \nl
Mauna Kea (Air Force)& 24" \nl
Mauna Kea (CFHT) & 3.6m \nl
McDonald & 30" \nl
McDonald & 36" \nl
McDonald & 82" \nl
OHP & 1.93m \nl
SAAO & 30" \nl
SAAO & 40" \nl
SAAO & 74" \nl
Siding Spring & 24" \nl
Siding Spring & 40" \nl

\enddata

\end{deluxetable}

\begin{deluxetable}{lllr}
\tablecaption{\label{tb:addjobs}Additional observations used in this work
that are not listed in Kleinman \etal (1994).}


\makeatletter
\def\jnl@aj{AJ}
\ifx\revtex@jnl\jnl@aj\let\tablebreak=\nl\fi
\makeatother



\tablewidth{33pc}
\tablehead{
\colhead{Telescope}           & \colhead{Run Name}      &
\colhead{Date}          & \colhead{Start}  \\
\colhead{} & \colhead{} & \colhead{(UT)} & \colhead{(UT)} }
\startdata
SAAO 74" & s3598 & 1985 Aug 8 & 0:54:04 \nl
SAAO 74" & s3606 & 1985 Aug 10 & 22:23:42 \nl
SAAO 30" & s3615 & 1985 Aug 13 & 20:56:20 \nl
SAAO 30" & s3618 & 1985 Aug 14 & 20:23:40 \nl
SAAO 30" & s3621 & 1985 Aug 15 & 20:24:00 \nl
SAAO 30" & s3624 & 1985 Aug 16 & 19:51:20 \nl
SAAO 30" & s3628 & 1985 Aug 17 & 20:14:20 \nl
SAAO 30" & s3631 & 1985 Aug 19 & 20:42:20 \nl
SAAO 40" & s3634 & 1985 Aug 20 & 21:05:16 \nl
SAAO 40" & s3638 & 1985 Aug 21 & 22:21:45 \nl
McDonald 36" & r3084 & 1985 Aug 22 & 6:37:26 \nl
McDonald 36" & r3085 & 1985 Aug 23 & 7:43:21 \nl
SAAO 40" & s3642 & 1985 Aug 23 & 20:54:18 \nl
SAAO 40" & s3645 & 1985 Aug 24 & 20:59:43 \nl
SAAO 40" & s3647 & 1985 Aug 25 & 20:06:20 \nl
McDonald 36" & r3086 & 1985 Aug 26 & 7:19:19 \nl
SAAO 40" & s3651 & 1985 Aug 26 & 20:20:03 \nl
SAAO 30" & s3654 & 1985 Sep 10 & 18:44:20 \nl
SAAO 30" & s3655 & 1985 Sep 13 & 17:42:00 \nl
SAAO 30" & s3656 & 1985 Sep 14 & 17:43:00 \nl
SAAO 30" & s3658 & 1985 Sep 15 & 0:42:40 \nl
SAAO 30" & s3660 & 1985 Sep 15 & 17:39:20 \nl
SAAO 30" & s3663 & 1985 Sep 16 & 17:39:20 \nl
McDonald 82"& r3088 & 1985 Oct 22 & 2:58:37 \nl
McDonald 82" & r3094 & 1985 Oct 31 & 3:02:00 \nl
McDonald 82" & r3095 & 1985 Nov 1 & 1:56:30 \nl
McDonald 36" & sjk-0264 & 1993 Jul 21 & 9:16:30 \nl
McDonald 36" & tkw-0034 & 1993 Aug 11 & 8:23:00 \nl
McDonald 36" & tkw-0040 & 1993 Aug 16 & 4:16:00 \nl
McDonald 36" & sjk-0265 & 1993 Sep 14 & 2:04:00 \nl
McDonald 36" & sjk-0266 & 1993 Sep 14 & 9:24:30 \nl
McDonald 36" & sjk-0267 & 1993 Sep 15 & 1:53:30 \nl
McDonald 36" & sjk-0268 & 1993 Sep 16 & 2:03:30 \nl
McDonald 36" & sjk-0269 & 1993 Sep 17 & 2:34:00 \nl
McDonald 82" & sjk-0270 & 1993 Sep 18 & 3:52:30 \nl
McDonald 82" & sjk-0276 & 1993 Sep 19 & 3:39:30 \nl
McDonald 82" & sjk-0277 & 1993 Sep 20 & 3:39:00 \nl
McDonald 82" & sjk-0278 & 1993 Sep 20 & 5:05:30 \nl
McDonald 82" & sjk-0279 & 1993 Sep 20 & 10:32:30 \nl
McDonald 82" & sjk-0281 & 1993 Sep 21 & 4:38:30 \nl
McDonald 36" & sjk-0282 & 1993 Nov 4 & 1:11:30 \nl
McDonald 36" & sjk-0283 & 1993 Nov 5 & 1:12:30 \nl
McDonald 36" & sjk-0285 & 1993 Nov 6 & 0:54:00 \nl
McDonald 36" & sjk-0287 & 1993 Nov 7 & 0:55:30 \nl
McDonald 36" & sjk-0288 & 1993 Nov 8 & 2:17:00 \nl
McDonald 36" & sjk-0289 & 1993 Nov 8 & 5:45:00 \nl
McDonald 36" & sjk-0293 & 1993 Dec 14 & 1:57:30 \nl
McDonald 36" & sjk-0295 & 1993 Dec 15 & 0:38:00 \nl
McDonald 36" & sjk-0297 & 1993 Dec 16 & 2:05:30 \nl
McDonald 36" & sjk-0299 & 1993 Dec 17 & 1:01:30 \nl
McDonald 36" & sjk-0302 & 1993 Dec 18 & 0:49:00 \nl
Siding Spring 24"& sjk-0363 & 1994 May 14 & 18:44:00 \nl
Siding Spring 24"& sjk-0368 & 1994 May 15 & 18:44:30 \nl
Siding Spring 24"& sjk-0373 & 1994 May 16 & 18:43:30 \nl
\enddata

\end{deluxetable}

The data were all reduced as described in Nather \etal (1990),
transforming the raw lightcurves to extinction corrected relative
intensity measurements.  The raw data consist of either two or three
lightcurves, measuring the variable star, the comparison star, and in
the case of three-channel instruments, the sky as well. With
two-channel instruments, we occasionally move the telescope off the target
and comparison stars to sample sky.  The instruments used for most of
these observations are described by Kleinman \etal (1996).

\section{Results}

\begin{figure}
\plotone{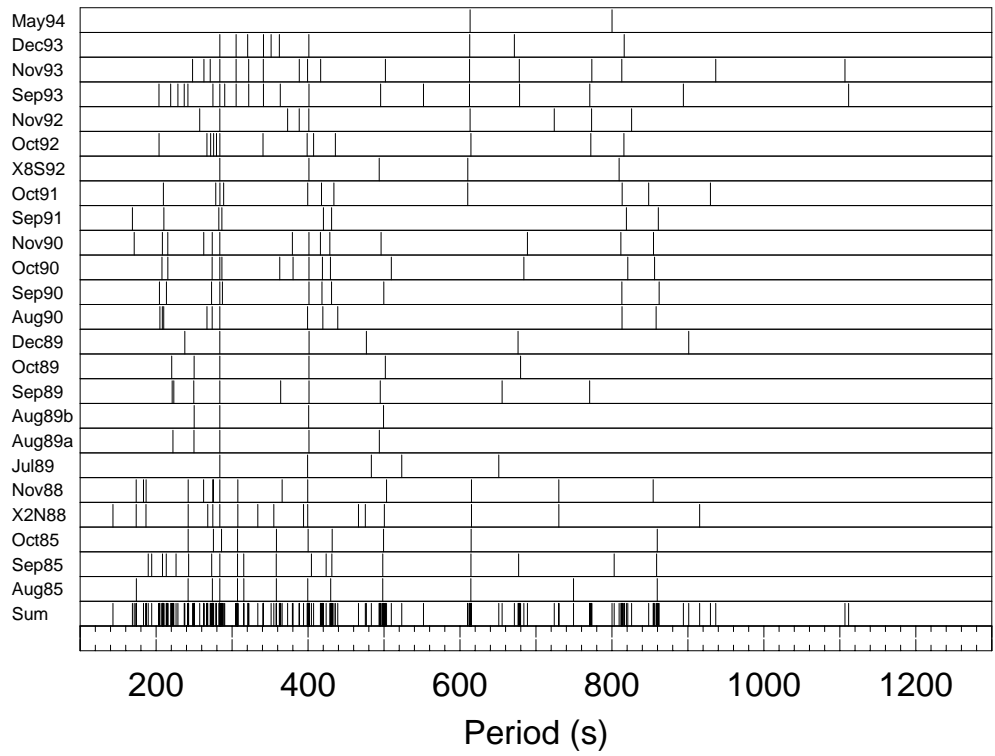}
\figcaption{\label{fig:g29peaks}Schematic diagram of G29--38's periodicities
for the entire data set.}
\end{figure}

After reducing each data set, typically a week or two in length, 
we calculated its FT and show them schematically in
Figure~\ref{fig:g29peaks}.  Each line represents a periodicity identified
at that period.  Since the dynamic range in the system
is large, we have ignored amplitude information and plotted every
periodicity with a line of the same height.  Each panel is labelled for
the month and year of the data, with the exception of X2N88 and X8S92,
which are data from the second WET run in November, 1988, and the
eighth WET run in September, 1992, respectively.  The bottom row of the plot,
labelled {\it Sum}, contains all the modes in the panels above,
collapsed into one.

While there is a wealth of information in this plot, the most striking
feature is the near continuum of modes shown in the {\it Sum} panel.
Were all these modes simply \l=1 and even \l=2 modes, the mode density
would not be nearly so great and there would be distinct gaps between
modes.

The most obvious explanation for this result is the known abundance of
linear combination frequencies which were not removed in the plot.  Our
ability to identify (and hence remove) the linear combinations depends
greatly on the \s2n ratio and the resolution of each transform.  To
help avoid uncertain and incorrect identifications, we now restrict the
analysis to the best data sets available --- one per year: Aug85, a
two-site campaign by South Africa and McDonald;  X2N88, the first WET
campaign;  Sep89, a 20-night data set from McDonald; Oct90 and Sep93,
slightly smaller single-site data sets from McDonald; and X8S92, the
second \g29 WET campaign.

Identifying the combination frequencies is a difficult task.  First, we
must identify the three frequencies that form the combination, then
decide which are the ``real modes'' and which is the combination.  That
is, if we find three signals with frequencies, A, B, and C such that
A+B=C, we need to decide if A is the difference frequency of two modes
B and C, if B is the difference frequency of A and C, or if C is the
sum frequency of A and B.  To identify the combinations, we wrote a
simple computer program that goes through a list of identified
frequencies from the power spectra.  In developing this code, we found
it identified every combination we found by hand, plus a few we did
not. In addition, while we often stopped after finding one possible
combination that contained a given mode, the computer code found all
possible combinations, often finding some more exact than those we had
originally identified.  The code takes into account possible
mis-identification of the frequencies if there are dominant aliases
present, and applies a selectable equality criteria in determining each
match.

Once we had identified the combinations, we decided which were real and
which were combinations using the following guidelines:  relative
mode amplitudes (combinations generally have smaller amplitudes),
number of combinations with each mode (combinations with combination
modes are less likely than first order combinations), the existence of
multiplet structure (two multiplets adding together produce distinctly
different structure in the combination), and the way modes and their
combinations disappear and recur in the various data sets (a {\it parent}
mode can appear without its combinations, but not vice versa). Still,
though, there were a couple occasions when we could not decide without resort
to a model that predicts normal modes.  In these cases, we assumed modes
with periods near the expected groups were real and not combinations.
We will gladly make available the list of frequencies or transforms to
anyone who wishes to seriously explore this procedure in more detail.
The FTs and mode lists are included in Kleinman (1995) as well.

\begin{figure}
\plotone{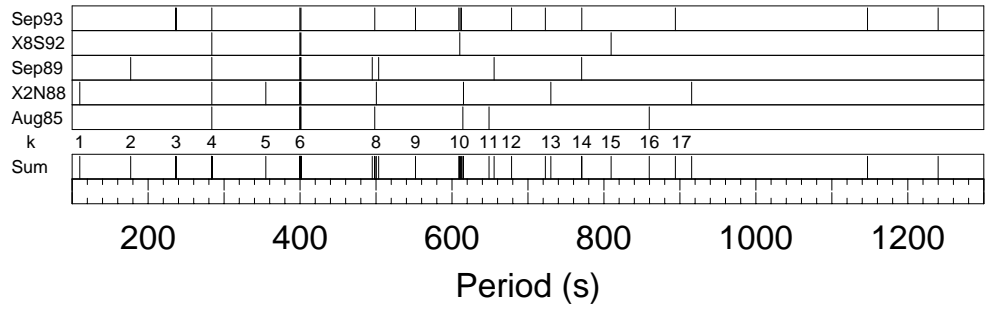}
\figcaption{\label{fig:g29peaks-lc}Schematic diagram of G29--38's
periodicities
minus the linear combination modes.}
\end{figure}

Figure~\ref{fig:g29peaks-lc} is the schematic period diagram for this
data subset, after removing the identified linear combination
frequencies.  This new schematic period diagram is substantially
cleaner and shows the mode groupings we would expect for
normal-mode pulsations.  The roughly equally-spaced groups seen in the
{\it Sum} row suggest a mean period spacing of roughly 50s, consistent with
models of $0.6\Msun$ DAVs \l=1 spacings (\cite{bra96}).  If these are
all \l=1 modes, we see a nearly complete set of them from 110s to 900s
with two more at longer periods.  With few exceptions, the groups are
tight and have distinct gaps between them.

\begin{figure}
\plotone{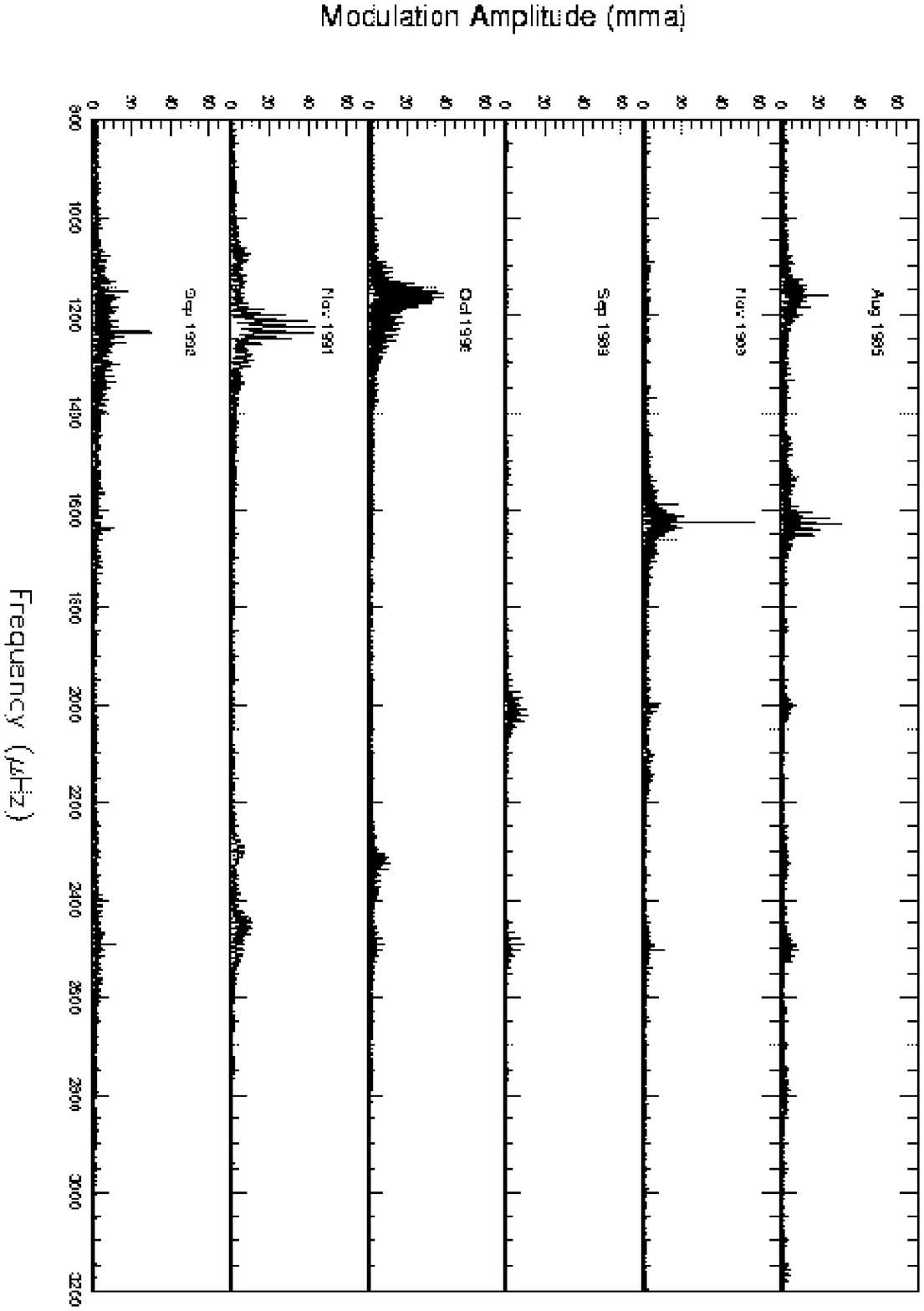}
\figcaption{\label{fig:g29fts}A portion of \g29's FT for several of our best
seasons. The large amplitude changes are obvious.}
\end{figure}

As Figure~\ref{fig:g29peaks} readily shows, the power spectrum of \g29
changes dramatically from year to year.  This is even more obvious in
Figure~\ref{fig:g29fts} which shows a portion of the FT, retaining real
amplitudes, for several of our best data sets. It appears, however, to
make its most dramatic changes when the star is behind the sun and we
cannot watch it. (This certainly sets an upper limit on the timescale
of change, something that might prove useful once we understand why it
makes such drastic changes.) While this predisposition means we do not
get to watch the star change, the advantage is that the FT of each
individual year provides a relatively stable set of modes that may be
used for \agl\ analysis.

\section{Mode Discussion}

As discussed earlier, we expect g-modes of identical \l\ values to be 
equally spaced in period in the asymptotic (high-\k) limit.
We cannot, however, now go and search for strictly uniform period spacing,
because we are {\it not} in asymptopia here.  We do not expect to find a
strictly uniform period spacing since we have identified modes possibly 
starting at \k=1, which is certainly {\it not} in the high-\k\ limit.
Instead, we expect to see significant departures until we get to the
higher-\k\ modes (our experiences shows \k=10 is a good rule of thumb).  We
also expect deviations on the order of 10s or so due to mode-trapping
effects.  In some cases, however, models of Bradley (1996) show deviations
of as much as 20--30s. 

We tried many different methods to quantitatively search for equal period
spacings in this observed set of modes while simultaneously allowing for the
inevitable departures from strictly uniform spacing, but did not succeed in
finding any significant results.  The fault may not, however, be in the data:
the same tests failed to find significant period spacings in some of the
model-derived test data as well.  We therefore believe that the inability
to statistically determine a significant mean period spacing does NOT
affect (either pro or con) the model of single-\l\ pulsations and
continue under the assumption that the approximate equal period spacing
seen in these modes is significant and that most likely, these are a series
of successive \k, \l=1 g-modes.  While the goal of studying this star, and
others like it, must be to uniquely determine the \k, \l, and \m\ for each
mode so we can proceed to explore the white dwarf interiors, were we to go no
further in the analysis than where we are now, we would be concluding with
the discovery of a a pattern to \g29's seemingly haphazard behavior: {\it a
recurring set of stable modes}. With its now understood wealth of normal
mode oscillations, \g29 has become the DAV of choice for \agy. 
We could now commence additional observing programs to help determine \l\
values including perhaps, the spectroscopic methods described by Robinson
\etal (1995).  In
addition, we now know how to make use of the entire set of
easy-to-observe, large-amplitude DA pulsators, to get a similar set of
information.  

However, since we are compelled by the observed period spacing, we
will continue the analysis under the assumption that each group is a
different \k, \l=1 mode. In figure~\ref{fig:g29peaks-lc}, we provide a
running \k\ assignment for each group. We will later discuss the
uncertainties of this assignment, but for now, they serve as references to
make the ensuing discussion easier.

The data from September, 1993 have the most modes and nicely reproduce
the $\approx$ 50s spacing seen in the {\it Sum} row earlier. This
season is unique in that there appear to be six consecutive overtones
(\k s) in one period range.  The other seasons also show roughly
equally spaced groups, but are often missing one or more \k s in
between each observed mode.  The {\it Sum} row shows all the gaps (but one)
are filled and we see what appears to be a set of same-\l\ modes from
\k=1 to \k=17.  Almost half of the observed modes
repeat at least once in the data set and four are present in four of the
five data sets.  We now have strong evidence for a series of successive-\k,
same-\l\ modes.

These same-\l\ modes fit well with an \l=1 model and its expected
period spacing with  the possible exceptions of the two modes
near \k=17,  although they could also be \k=17 and \k=18 without much
problem.  These two modes, the 894s mode in Sep93 and the 915s mode in
X2N88, are separated by 21s, a little too close, but not completely
impossible, to be different \k s (17 and 18), same-\l\ pulled closer by
mode trapping.  Models of Brassard \etal (1992) and Bradley
(1993) with hydrogen layer masses near $10^{-10}\Mstar$ do, for
example, show the strong mode trapping this identification implies near
these periods, although this match requires a rather limited parameter set.
In frequency, the separation is 26 \mhz, too large to
easily fit the frequency spacings of the known multiplets.  We cannot
yet completely settle this ambiguity, but are relieved to know it does
not affect the rest of the pattern.

\subsection{Identification Ambiguities}

As the two footnotes in our list of observed periods, Table~\ref{tb:modes},
suggest, there are still
some ambiguities in our mode identification assignments.   The two
most critical uncertainties, as luck would have it, are for the two
modes most critical in model-matching: the lowest-\l\ modes.  The \k=1 mode
seen in the X2N88 data set has two combinations: plus and minus the
\k=10 mode.  There are no other combinations with these modes, we don't
see them in any other year, and their amplitudes are all quite similar,
thus it is impossible with this data to determine which mode is not a
combination mode.  The most likely choice is that we are seeing
the \k=1 mode at 110s and its sum and difference with the \k=10 mode.
However, we could also be seeing the \k=1 mode at 134s with two sums
with the \k=10 mode.  It is unlikely, given the preference for observable
sums over differences, that both the 110s and 134s are differences with
the higher frequency \k=1 mode at 93s.

The \k=2 mode identification seen in the Sep89 data set is also uncertain,
but this time not due to the presence of combinations, but rather the lack
of them. This mode is of extremely low-amplitude and is not well-resolved in
the FT.  Because of the very high-quality data in this data set, the power
seen in the region is probably significant, but we cannot say for certain
where the mode is or even if it is stable over the course of the run.
Like the \k=1 mode, we don't see this mode in any other data set.
Placing the mode at the highest peak in the region and identifying it as
\l=2 fits in well with the expected mode pattern, but is highly uncertain.

Also in the X2N88 data is a messy region of excess power near
2105\mhz, its largest-amplitude component. We also see a series of
three combination sums of this region (and the 2105\mhz\ peak) with  the
large-amplitude \k=10 mode.  One of either the 2105\mhz\ peak, or its
sums with the \k=10 mode, therefore, must be a  real oscillation mode,
but we cannot decide which it is.  No matter which mode it is, however,
it does not appear to fit in well with the established pattern of \l=1
modes and their associated period spacings.

In the Sep89 data, we see two modes at 1986 and 2020 \mhz, near what we
have been calling the \k=8 region, but also significantly different from
the power seen there in other years. As seen in the next section, this
mode shows strange multiplet structure and the modes seen here could
be some artifact of that. Or, they could be something else entirely;
we cannot yet say.

Also in the Sep89 data are two modes at 2747\mhz\ and its sum with the
1986\mhz\ mode.  The 2747\mhz\ mode is about 3 times larger in amplitude
than the sum mode, and thus is likely the non-combination mode, but we
have no other evidence to support this.  Neither mode fits in well with
our \l=1 spacings.

We suspect therefore, that we do have some non-\l=1 modes in our data,
but the lack of repeatability and consistently low-amplitude makes it
difficult to say with certainty.  If we accept the \l=1 assignments for
the \k=1, \k=2, and perhaps even the \k=8 modes (1986 and 2020 \mhz)
of Sep89,  then we have only two modes, the 2105\mhz\ group of X2N88 and
the 2747\mhz\ pair of Sep89 to explain some other way or with some other
\l.

\section{Multiplets}

\begin{figure}
\plotone{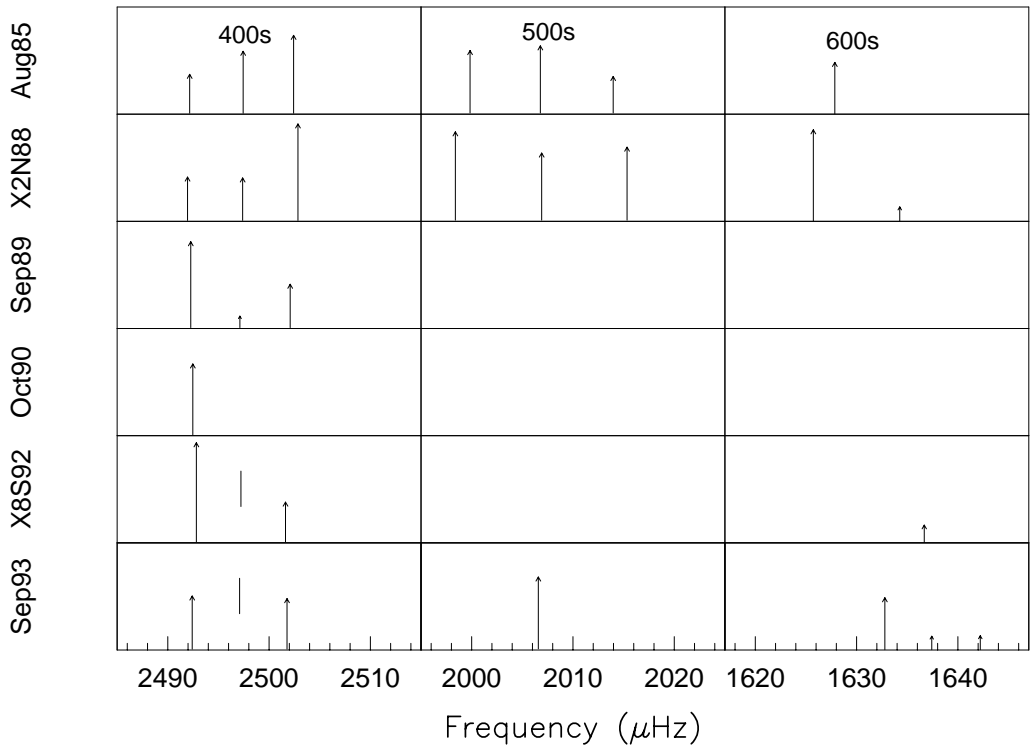}
\figcaption{\label{fig:allmulti}A schematic plot of all the major observed
multiplets.  The short line segments are not real modes, but show the
average period of the two flanking modes.}
\end{figure}

While we have uncovered strong suggestions for the non-combination
modes being all \l=1, we would like additional proof.  Ordinarily,
rotational splitting of different \m\ values for a given \l\ and \k\ is
a valuable aid in assigning \l-values.  Unfortunately, we do not find
many multiplets in this data set.  Those that are present do not behave
quite as theory predicts.  Since a good picture is often worth a few
pages of text, we have schematically plotted in
Figure~\ref{fig:allmulti} the resolved multiplets.  This time, the
height of each arrow is proportional to the amplitude of the mode
(although the scale changes for each multiplet series) and the shorter
line segments are not real modes, but simply represent the average
period of the two modes on either side of it and is meant to suggest
the location of the presumed missing \m=0 mode.

Only twice do we see identical spacings within a given data set.
In the X2N88 data, the 500s and 615s multiplets both have a $\approx$
8.5\mhz\ spacing.  In the Sep93 data, the 400s and 612s multiplets both
have an $\approx$ 4.7\mhz\ spacing.
The 400s multiplet always has the
smaller spacing; the 500s multiplet, the larger; and the 600s modes
alternately fall into both camps.  
Also note the 400s multiplet spacing
is not constant, but varies (perhaps periodically) with time.  All
these results are quite interesting, and likely to be clues to the
complicated structures in this star, but seem to be largely irrelevant
to the overall mode picture. That is not to say there is no information in
these changes, however; they may very well be understood by some nonlinear
effects such as those discussed by Buchler, Goupil, and Serre (1995). We 
have only recently begun to acquire the kind of data that could comfortably
test such theories.  The unfortunate thing here is they do not directly yield
information about the \l-value of our modes, but seem to be even more
mysterious if we invoke multiple-\l s than if we don't.  Thus they support,
but do not demand, our \l=1 model.

\section{Analysis}

We have now discovered a fairly regular pattern of recurring modes in the
power spectra of \g29 and have suggested they may all form a pattern of
successive \k, \l=1 g-modes.  Going just this far is a very important step
to uncovering the \agl secrets hidden inside these stars; we never before
knew if the modes we were seeing were sustained, normal-modes of oscillation 
or something much more fleeting and chaotic.  This is just the first step,
however.  What remains now is to explore models of DAVs, in attempts to tune
the input physics to the observed set of frequencies.  Such modelling work
is a large task and is best allotted a separate work for
the detailed, careful analysis and discussion which must inevitably occur.
However, here we can address  the direction
such an effort might take.  We do not intend these results to
be concrete, but they should help point the modelling investigation
in a productive direction.

The goals of the detailed investigation will, of course, be to
constrain or determine the mass and internal structure of \g29.
The asteroseismologically-determined estimates of both the total stellar
and hydrogen layer masses will be particularly important in helping to 
both calibrate the \agl methods with other mass determinations and in 
checking models of white dwarf formation and evolution which set limits on
the amount of hydrogen in the DAVs.  We also want to see how \g29
fits in with the other DAVs, both the hot and cool ones:  does it fit
the period structure of the hot DAVs found by Clemens (1993, 1994) and
does the period spectra of \g29 look like that of the other cool DAVs,
to the extent we can tell?  We must defer the full answers to these questions
to papers now in progress (however, see \cite{kle95}, \cite{dol97}), but
will address some of the key issues in a preliminary way here.

\begin{figure}
\plotone{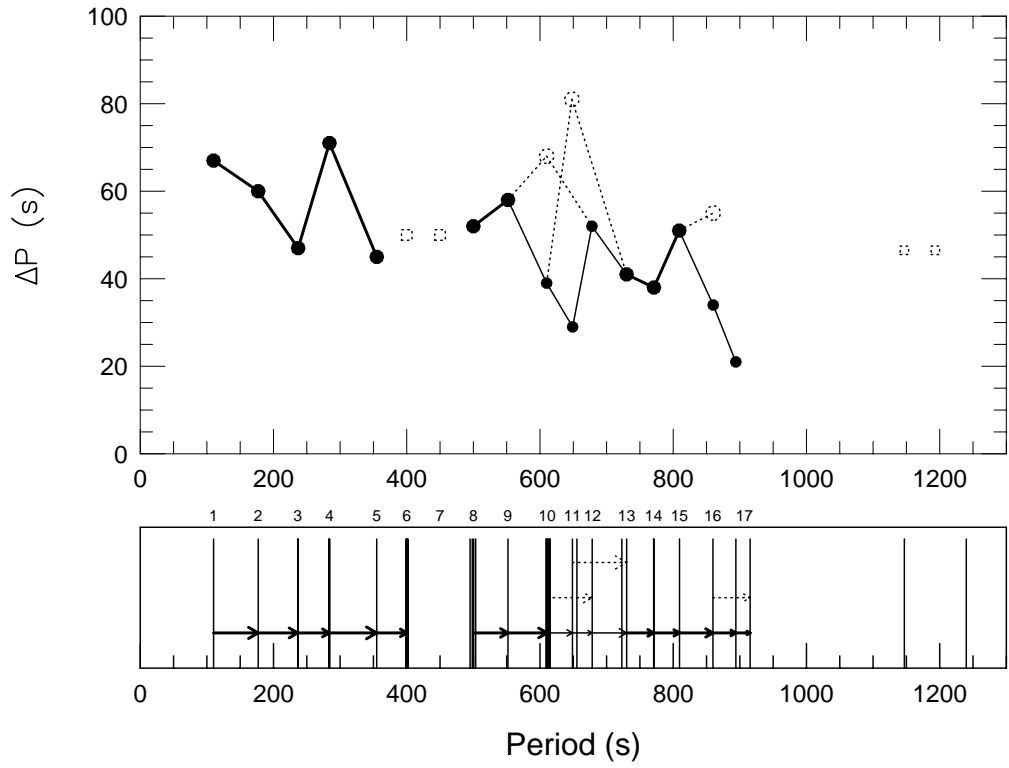}
\figcaption{\label{fig:dpvsp}The observed $\Delta$P vs. P diagram for
G29--38.}
\end{figure}

If this observed set of modes is a set of single-\l\ pulsations, then
the mean period is an indicator of mass and the deviations from the
mean describe the layered internal structure of the star.  Using
forward differences (thus defining $\Delta$P), we have plotted in
Figure~\ref{fig:dpvsp} a $\Delta$P vs. P, or period spacing, diagram for
these data.  With the possible exception of the two modes near \k=17, there
is no reason to suspect any of these modes are a different \l\ (we later
treat these two \k=17 modes as \k=17 and \k=18).  Based on the evidence
already discussed and previous \agy from
WET observations and the work of Clemens with the hot DAVs, it
seems unlikely that the majority of the modes are anything other than \l=1.
However, since it is possible we have inadvertently included a mode or two
of different \l\ into our identifications, we have plotted some likely
alternatives to the the \l=1 only model by dotted lines and circles. Where
we have plotted such alternatives, the only \l=1 interpretation is plotted
with thinner lines and slightly smaller filled circles. The dotted squares
represent the average spacing for two modes with an assumed missing
mode between them.  The bottom panel of the plot is an enlarged plot of the
sum row of Figure~\ref{fig:g29peaks-lc} with arrows showing which modes
were used to calculate the differences shown above it. Above this box are
\l=1 \k\ assignments. Each point has an additional uncertainty of a few
seconds since we cannot be certain of the \m\ values. In the worst case,
the shift can be as much as 14s for periods near 900s, and 3s for periods
around 400s, assuming the largest splitting of 8.5\mhz. With the smaller
4.7\mhz\ splitting, the worst case is about 8s. Table~\ref{tb:modes} has a
list of the periods used in this diagram (minus multiplets and repeated
modes) along with the \k\ assignment in the only \l=1 model.

\begin{deluxetable}{lr}
\tablecaption{\label{tb:modes}Observed periods used in Figure~\protect\ref{fig:dpvsp}. The
\k\ assignments assume all the modes are \l=1.}


\makeatletter
\def\jnl@aj{AJ}
\ifx\revtex@jnl\jnl@aj\let\tablebreak=\nl\fi
\makeatother

\tablewidth{18pc}
\tablehead{
\colhead{\k}           & \colhead{Period}\nl
\colhead{}             & \colhead{(s)}}
\startdata
1\tablenotemark{a} & 110\nl
2\tablenotemark{b} & 177\nl
3 & 237\nl
4 & 284\nl
5 & 355\nl
6 & 400\nl
8 & 500\nl
9 & 552\nl
10 & 610\nl
11 & 649\nl
12 & 678\nl
13 & 730\nl
14 & 771\nl
15 & 809\nl
16 & 860\nl
17 & 894\nl
\tablebreak
18 & 915\nl
?? & 1147\nl
?? & 1240\nl
\tablenotetext{a}{This mode has a sum and difference combination with the
k=10 mode, thus the ``real'' non-combination mode is either 110s as shown
here, or 134s.}
\tablenotetext{b}{The identity of this mode is questionable.  There is
excess power in this region, but isolating it to a particular
frequency is difficult. It should not be a strong constraint in
model-fitting attempts.}
\enddata

\end{deluxetable}

Current published models (Bradley 1993, 1995) do not do a very good job of
fitting this diagram, but we are encouraged by this as it means there
is something we can learn from new modelling efforts.  What is
different about this diagram is the trend, or perhaps sudden change, to
lower period spacing with higher period.  This trend could be revealing
physics unaccounted for in the model star, or it could mean some of
these later modes are not \l=1, as some of the alternative
identifications plotted in Figure~\ref{fig:dpvsp} suggest.  Taking the
dotted line path, this trend effectively disappears, but we have to
come up with new assignments for the skipped-over modes.  Bradley \&
Kleinman (1997) have shown, however, they can get a reasonable DA model to
match the observed modes assuming mostly \l=1.

Based on the \l=1 only interpretation of the observed modes, we
calculate a mean period spacing, $P_\circ$, of $47\pm12$s.  This value
can increase by a few (3--4) seconds, with the standard deviation
decreasing an almost equal amount, if we take some of the alternative
identifications.  For now, we will work with the 47s spacing,
determined by assuming {\it all} the modes are the \l=1 and have the
\k\ assignments made earlier.  (We have here called the two modes near
\k=17, \k=17 and \k=18.)

Using models of Bradley (1996) with a hydrogen layer mass of $1.0
\times 10^{-4}\Mstar$, consistent with the results of Clemens
(1993, 1994), and a temperature
of 11820~K from Bergeron \etal (1995), this period spacing corresponds
to a mass of roughly $0.60\Msun$, quite in line with the mean observed
white dwarf mass.  For this comparison, we used the 11700~K,
0.6$\Msun$, standard model with $1.0\times 10^{-4}\Mstar$ hydrogen
layer and helium layer of $1.0 \times 10^{-2}\Mstar$, of Bradley (1996)
with its mean \l=1 period spacing (determined from \k=1 to \k=15) of
$47 \pm 14$~s.  Unfortunately, this particular model does not match
the individual modes very well, but it helps to establish the observed
period spacing is in the correct range for a reasonable choice of
parameters and to note the standard deviations of the model spacings and
our observed spacing are similar.

The mean period spacing is mainly a function of overall stellar mass.
The hydrogen layer mass, however, also has an effect, and once we allow
it to be a free parameter, we must consider its effects when determining
a model \agl fit.  If we have another mass measurement, we can fix the
overall stellar mass and just vary the hydrogen layer mass to match the
observed mean period spacing.  The most recent \g29 mass estimate is
from Bergeron \etal (1995).  They derive a value near 0.69~$\Msun$.  We
can fit this mass with our $P_\circ$ if we use a $10^{-10}\Mstar$
hydrogen layer mass (\cite{bra93}), although once again, the detailed
mode list doesn't match very well.

The most critical mismatch of the above-mentioned models to our list of
modes are the \k=1 and \k=2 modes.  The $0.6\Msun$ model, for example,
has its \k=1, \l=1 mode at 145~s; the $0.7\Msun$ model has its
\k=1, \l=1 mode at 210s. Our labelled \k=1 and \k=2 modes are at 110s and
177s respectively.  There is clearly a mismatch here. The observed \k=1 and
\k=2 modes are certainly the lowest amplitude of the other modes, and
therefore more prone to uncertainty, and could possibly be spurious.  The
case against this, however is fairly strong, as all the frequencies
identified with lower amplitudes than these turned out to be linear
combination frequencies; hence their amplitudes are indeed significant.  It
could also be these modes are part of unidentified combinations, but there
is as yet no evidence for this.  It would be nice to see these modes appear
again in additional (perhaps archival) data to help answer these
questions.  Given the ambiguities in the identification of these two modes
already discussed, modelling efforts will have to be very careful and
rigorous with all the possibilities here.  

There is one thing we can say with a much greater confidence:  the
identified set of non-combination peaks are {\it not} predominantly
\l=2 modes.  Such a spacing with \l=2 modes implies a stellar mass near
$0.2\Mstar$, well out of the range of all previous mass estimates. The
majority of the modes must be \l=1.

Preliminary results (\cite{kle95}, \cite{dol97}) show that where modes
have been identified in other cool DAVs (from a slightly less comprehensive
data set than that presented here), there are usually modes in \g29 as well.
This result extends that of Clemens to include the cooler DAVs and will be
published in a future paper.  If we follow the same line of reasoning as
Clemens, this uniformity also suggests we have identified predominantly
\l=1 modes.

\section{Conclusions}

After many years of searching, we finally have a DAV, \g29, with enough
observed modes to make a detailed \agl analysis possible.  With the
extensive, 10-year data set presented here, we have separated the
linear combination frequencies from the normal modes that recur in the
star's Fourier spectrum.  We have uncovered a discrete set of modes
from 110s to 1193s with only a few gaps.  We present evidence to
support the assertion that these modes are predominantly \l=1
pulsations. There have never been so many modes in a DAV offered for
analysis; we therefore expect surprises.  Preliminary modelling efforts
do indicate we can match these modes with current models, but so far
the matches are not quite as close as we would prefer. These efforts are only
preliminary as we have not yet attempted a systematic search of
parameter space to fit these observations.

Depending on the exact value of the $P_\circ$, the period spacing, and
the mass of the hydrogen layer, there may be a discrepancy between the
\agl mass and other measurements.  The solution to this potential
disagreement is as yet unknown, but perhaps more careful modelling
efforts will guide us to the answer: are our mode identifications
incorrect, the models lacking, or is \g29's behavior affecting the
other mass-determination methods, biasing them to an incorrect answer?
Undiscussed since the Introduction, as it seemed to be
completely independent to the presentation up to this point, however,
is \g29's infrared excess.  Clearly something strange is happening to
this star and until we definitively tie down the cause of the infrared
excess, we cannot rule out its possible effects on the pulsations.  We
are encouraged, however, that Zuckerman (private communication) reports
\g29's IR excess remains unique among the DAVs and that so far, \g29's
pulsation spectrum looks similar to other cool DAVs.  An intriguing
possibility, however, involves the effect of the large amplitude
pulsations on an H-layer close to the nuclear-burning limit.  Could this
provide the source for both the pulsation instabilities and the IR
excess? Koester \etal (1997) suggest not. They report atmospheric detection
of heavy elements on the surface of \g29, consistent with an increased
accretion rate from surrounding circumstellar material, perhaps settling
the mystery of \g29's infrared excess.

To help address these questions, we have obtained data  on many of
the other cool DA pulsators.  Preliminary results
do not yet address the mass problem (as not enough modes have been
identified), but the modes we have seen all appear nearly identical to
those in \g29.  This result, when added to Clemens's earlier work on
the hotter DAVs, really says the DAs are a very homogeneous
class.  Therefore, given this set of modes observed in \g29, it is
extremely unlikely that they are anything but \l=1 pulsations such as
have been found in the hotter DAVs.   To summarize in broad terms, then,
the evidence presented here for a set of \l=1 modes with a near 47s
spacing in \g29 is the following:

\begin{itemize}
\item There is a fairly regular $\approx$47s period spacing observed.   This value fits
well in the range expected for stars with masses between 0.5 and 0.7$\Msun$.
\item The mean period spacing is too high to be \l=2 and maintain a reasonable mass for \g29.
\item When we see multiplets, we see only triplets, consistent with only \l=1
rotationally-split modes.
\item The cool DAVs as a class share many common modes which appear to overlap those
of the hotter DAVs which have already been suggested are \l=1 modes
(\cite{kle95} and \cite{dol97}). 
\end{itemize}

If we are going to label these modes as something other than
\l=1 g-modes, we have to explain these four observations as
coincidences.  One accidental agreement is relatively easy to dismiss;
four are much more difficult.  There is no easy way to explain the majority
of these modes as anything {\it but} \l=1.
 
The suggestion of global uniformity in the DA stars is so profound, we
must look at it more closely. In particular, we must direct our efforts
at determining both the stellar and hydrogen layer masses.  We are now
continuing to observe other DAs to determine the extent of their
homogeneity. In  February, 1996, the WET targeted the cool DAV,
HL~Tau~76.  We hope data from this run, plus previous single-site runs,
will uncover a similarly rich pulsation spectrum as we found here.
Hopefully, the results form these observations will help guide us to
some of the answers we posed earlier.   These results should also help
address the source of our model fitting problems, should they remain
after a more detailed attempt.

No matter what the numerical results of future modelling are, we have
now made the necessary first step for their analysis: we have shown the
cool DAVs are normal-mode pulsators whose stable modes reappear in
precise places in the power spectra, amidst a veritable forest of
combination modes.  We have taken a star whose power spectra seemed
unsolvable, rapidly varying, and incapable of providing the kinds of
clues needed for an \agl analysis, and extracted such very clues.  We
must now, not only carefully explore model space to match these
observations, but obtain similar data on other stars similar to \g29 so
we can begin to determine which conflicts between star and model belong
only to \g29, and which belong to the entire group of stars and
therefore, to  our understanding  of the general physics which we put
into the models.  An increased understanding of the physical nature and 
evolution of the white dwarfs and the chance to calibrate more
accurately the white dwarf cooling sequence, and hence the age of the
local galactic disk, await only models to fit to these and future
observations.

\acknowledgments

S.J.K. wishes to extend his thanks to E. Presley for his suggestions on 
an appropriate use of black velvet.  This work was supported in part by
NSF Grants AST-8552457, AST-8600507,  AST-9013978, AST-9014655, AST-9217988,
AST-9314803, INT-93-14820,
and National Committee for Scientific Research Grant
2-2109-91-02 (Poland).  G.F. wishes to 
acknowledge the support of the NSREC of Canada and that of the fund FCAR
(Quebec).

\clearpage

\end{document}